\documentclass[allclo]{FBSart}
\usepackage{amsfonts}
\usepackage{amssymb}
\usepackage{graphicx}

\title{Scattering Length for Helium Atom-Diatom Collision\thanks{Contribution
to Proceedings of the International Workshop ``Critical
Stability of Few-Body Quantum Systems'' (Dresden, October 17--22,
2005). This work was supported by the Deu\-t\-s\-che
For\-s\-ch\-ungs\-ge\-mein\-schaft (DFG), the Heisenberg-Landau
Program, and the Russian Foundation for Basic Research.}}
\author{ E. A. Kolganova\instnr{1,2},
A. K. Motovilov\instnr{2}, and W. Sandhas\instnr{3}}
\instlist{Max-Planck-Institut f\"ur Physik komplexer Systeme,
N\"othnitzer str. 38, 01187 Dresden, Germany \and
Bogoliubov Laboratory of Theoretical Physics,
JINR, Joliot-Curie 6, 141980 Dubna, Moscow Region, Russia
\and Physikalisches Institut, Universit\"at Bonn,
Endenicher Allee 11-13, D-53115 Bonn, Germany }

\runningauthor{ E. A. Kolganova,
A. K. Motovilov, and W. Sandhas}
\runningtitle{Scattering Length for Helium Atom-Diatom Collision}
\begin{document}

\maketitle
\begin{abstract}
We present results on the scattering lengths of
$^4$He--$^4$He$_2$ and $^3$He--$^4$He$_2$ collisions.
We also study the consequence of varying the coupling constant of the
atom-atom interaction.
\end{abstract}

\section{Introduction}

The two-body scattering length in a dilute gas of alkali atoms can
be varied by changing the external magnetic field close to a
Feshbach resonance \cite{Duine}. In this way one may force the
two-body s-wave scattering length to go from positive to negative
values through infinity. Therefore, the magnetic field should be an
appropriate tool in modeling the Efimov effect. We recall that this
effect occurs in case of infinite two-body scattering lengths,
manifesting itself in an infinite number of three-body bound states.

This role of the magnetic field combined with a Feshbach resonance
may be mimicked by varying the coupling constant of the two-body
interaction within a three-body system that is not necessarily
subject to a magnetic field \cite{Esry}. In this context the system
of three $^4$He atoms appears to be the best candidate. Actually, it
has been shown that the excited state of the $^4$He trimer is
already of Efimov nature. To get the complete Efimov effect it
suffices to weaken the He--He interatomic potential only by about
3\%.

In the present work we extend the investigation of the
three-atomic helium systems undertaken in 
\cite{MSSK}, which was based on a mathematically rigorous hard-core version of
the Faddeev differential equations. We calculate
the scattering lengths for $^{3;4}$He atoms - $^4$He dimer collisions. Under the
assumption that weakening the potential mimics the behaviour of the scattering length
in a magnetic field, we show the dependence of low-energy three-body scattering
properties on the two-body scattering length.

Some of the results presented in this paper were reported already
in~\cite{PRA2004} and~\cite{FBS2004}.

\section{Results}

In our calculations we employed the hard-core version of the Faddeev
differential equations developed in \cite{MSSK}. As
He-He interaction we used the semi-empirical HFD-B \cite{Aziz87}
and LM2M2 \cite{Aziz91} potentials by Aziz and co-workers, and the
more recent, purely theoretically derived TTY \cite{TTY} potential
by Tang, Toennies and Yiu. For the explicit form of these
polarization potentials we refer to the Appendix of Ref.
\cite{MSSK}. As in our previous calculations we choose
$\hbar^2/m_{_{^4{\rm He}}}=12.12$\,K\,\AA$^2$ and $m_{_{^3{\rm
He}}}/m_{_{^4{\rm He}}}=0.753517$ where $m_{_{^3{\rm He}}}$ and
$m_{_{^4{\rm He}}}$ stand for the masses of the $^3$He and $^4$He
atoms, respectively. The $^4$He dimer binding energies and
$^4$He--$^4$He scattering lengths obtained with the HFD-B, LM2M2,
and TTY potentials are shown in Table \ref{tableDimerLen}. Note that
the inverse of the wave number $\varkappa^{(2)}=\sqrt{|\epsilon_d|}$
lies rather close to the corresponding scattering length.


\begin{table}[hb]
\caption{Dimer energy $\epsilon_d$, wave
length $1/\varkappa^{(2)}$, and
$^4$He$-$$^4$He scattering length $\ell_{\rm sc}^{(1+1)}$
for the potentials used, as compared to the experimental
values of Ref. \cite{exp}.} \label{tableDimerLen}
\begin{center}
\begin{tabular}{ccc|cccc}
\hline
 &  $\epsilon_d$ (mK) &  $\ell^{(1+1)}_{\rm sc}$ (\AA) &
{ Potential} &  $\epsilon_d$ (mK) &  $1/\varkappa^{(2)}$ (\AA)
&  $\ell^{(1+1)}_{\rm sc}$ (\AA)  \\
\hline
  &   &  &  LM2M2 & $-1.30348$ & 96.43 & 100.23 \\
 Exp. \cite{exp} & $1.1^{+0.3}_{-0.2}$  &$104^{+8}_{-18}$ & TTY   & $-1.30962$ & 96.20 & 100.01 \\
 &   & & HFD-B  & $-1.68541$ &  84.80 & $ 88.50$ \\
\hline
\end{tabular}
\label{T2body}
\end{center}
\end{table}
\vspace*{-0.5cm}
\begin{table}[htb]
\label{tableScLength}
\caption{The $^{4}$He--$^4$He$_2$ scattering length $\ell^{(1+2)}_{\rm sc}$
 ({\AA})
on a grid with
$N_\rho=N_\theta$=2005 and $\rho_{\rm max}$=700 \AA. }

\begin{center}
\begin{tabular}{ccccccccc}
\hline
{ Potential}  & { $l_{\rm max}$} &
{ This work }& { \cite{MSSK}}  & { \cite{BlumeGreene}}
& { \cite{RoudnevE}} & {\cite{Penkov} } &
{ \cite{BraatenHammer} } & \\
\cline{1-2}\cline{3-9}
       &  0 & 158.2& $168$  &     &       &&& \\
 LM2M2 &  2 & 122.9 & $134$  &     &       &&& \\
       &  4 & 118.7 & $131$  &  126  & 115.4 &114.25&113.1& \\
\cline{1-2}\cline{3-9}
       &  0 & 158.6& $168$   &     &        &&&  \\
 TTY   &  2 & 123.2& $134$   &     &        &&& \\
       &  4 & 118.9& $131$   &     & 115.8  &&114.5& \\
\cline{1-2}\cline{3-9}
       &  0 & 159.6& $168$   &     &        &&& \\
 HFD-B &  2 & 128.4& $138$   &     &        &&& \\
       &  4 & 124.7& $135$   &     & 121.9  &&120.2& \\
\hline
\end{tabular}
\end{center}
\end{table}

We have improved our previous calculations~\cite{MSSK} of the
scattering length by increasing the values of the grid parameters and
cutoff hyperradius. The corresponding results are presented in Table
\ref{tableScLength}. This table also contains the fairly recent
results by Blume and Greene \cite{BlumeGreene} and by Roudnev
\cite{RoudnevE}. The treatment of \cite{BlumeGreene} is based on a
combination of the Monte Carlo method and the hyperspherical
adiabatic approach. The one of Ref. \cite{RoudnevE} employs the
three-dimensional Faddeev differential equations in the total
angular momentum representation. Our results agree rather well with
these alternative calculations.

For completeness we mention that besides the above {\em
ab initio} calculations there are also model
calculations, the results of which are given in the
last two columns of Table \ref{tableScLength}. The
calculations of \cite{Penkov} are based on employing a
Yamaguchi potential that leads to an easily solvable
one-dimensional integral equation in momentum space.
The approach of \cite{BraatenHammer} represents
intrinsically a zero-range model with a cut-off
introduced to make the resulting one-dimensional
Skornyakov-Ter-Martirosian equation \cite{STM} well
defined. The cut-off parameter in
\cite{BraatenHammer} as well as the range
parameter of the Yamaguchi potential in \cite{Penkov}
are adjusted to the three-body binding energy obtained
in the {\it ab initio} calculations. In other words, these
approaches are characterized by remarkable
simplicity, but rely essentially on results of the {\it
ab initio} three-body calculations.

\begin{table}[htb]
\caption{The $^{3}$He--$^4$He$_2$
 atom-dimer scattering length $\ell^{(1+2)}_{\rm sc}$(in \AA).}
\label{tScLength3}
\begin{center}
\begin{tabular}{ccccccccc}
\hline
Potential & \multicolumn{3}{c}{LM2M2}&&\multicolumn{3}{c}{ TTY} \\
\cline{1-2}\cline{3-5}\cline{7-9}
\qquad $l_{\rm max}$\quad && 0 & 2 & 4 & & 0 & 2 & 4 \\
\cline{1-2}\cline{3-5}\cline{7-9}
This work && 38.5 & 22.2 & 21.0 & & 38.8 & 22.4 & 21.2 \\
\cite{Roudnev} & & & & 19.3 & & & & 19.6\\
\hline
\end{tabular}
\end{center}
\end{table}
\vspace*{-0.5cm}
\begin{table}[htb]
\caption{Dependence of the trimer energies (mK) and the scattering
length (\AA) on the potential strength $\lambda$ (for $l_{\rm
max}=0$). } \label{tableScLength1}
\begin{tabular}{lccccccc}
\hline $\lambda$ & $\epsilon_d$ & $\epsilon_d - E^{(1)}_{\rm ex}$ &
$\epsilon_d - E_{\rm virt}$ & $\epsilon_d - E^{(2)}_{\rm ex}$ &
$\ell^{(1+2)}_{\rm sc}$ & $\ell^{(1+1)}_{\rm sc}$ & $\rho_{\rm max} (\AA)$ \\
\hline
1.0 & -1.685 & 0.773 & - & - & 160 & 88.6 & 700 \\
0.995 & -1.160 & 0.710 & - & - & 151 & 106 & 900 \\
0.990 & -0.732 & 0.622 & - & - & 143 & 132 & 1050\\
0.9875$\quad$ & -0.555 & 0.222 & - & - & 125 & 151 & 1200 \\
0.985 & -0.402 & 0.518 & 0.097 & - & 69 & 177 & 1300 \\
0.982 & -0.251 & 0.447 & 0.022 & - & -75 & 223 & 1700 \\
0.980 & -0.170 & 0.396 & 0.009 & - & -337 & 271 & 2000\\
0.9775 & -0.091 & 0.328 & 0.003    & - & -6972 & 370 & 3000\\
0.975 & -0.036 & 0.259 & - & 0.002 & 7120 & 583 & 4500\\
0.973 & -0.010 & 0.204 & - & 0.006 & 4260 & 1092 & 10000\\
\hline
\end{tabular}

\end{table}

Due to the smaller mass of the $^3$He atom, the $^3$He -- $^4$He
system is unbound. Nevertheless, the $^3$He$^4$He$_2$ trimer exists,
though with a binding energy of about 14 mK (see \cite{FBS2004} and
references therein). And, in contrast to the symmetric case, there
is no excited (Efimov-type) state in the asymmetric $^3$He$^4$He$_2$
system. Table~\ref{tScLength3} contains our results for the
$^3$He--$^4$He$_2$ scattering length.

Following the idea that weakening the potential could imitate the
action of a magnetic field on the scattering length, we multiply the
original potential $V_{\rm HFD-B} (x)$ by a factor $\lambda$.
Decreasing this coupling constant, there emerges a virtual state of
energy $E_{\rm virt}$ on the second energy sheet. This energy,
relative to the two-body binding energy $\epsilon_d$, is given in
column 4 of Table \ref{tableScLength1}. When decreasing $\lambda$
further, this state turns into the second excited state. Its energy
$E^{(2)}_{\rm ex}$ relative to $\epsilon_d$ is shown in the next
column. These energy results are in a good agreement with the
literature \cite{EsryLinGreene}. When the second excited state
emerges, the $^{4}$He--$^4$He$_2$ scattering length
$\ell^{(1+2)}_{\rm sc}$ changes its sign going through a pole, while
the two-body scattering length $\ell^{(1+1)}_{\rm sc}$ increases
monotonically.

\begin{acknowledge}
We are grateful to Prof.~V.\,B.\,Belyaev and Prof.~H.\,Toki for providing us
with the possibility to perform calculations at the supercomputer of the
Research Center for Nuclear Physics of Osaka University, Japan. One of us
(E.A.K.) is indebted to Prof. J. M. Rost for his hospitality at the
Max-Planck-Institut f\"ur Physik komplexer Systeme, Dresden.
\end{acknowledge}

\end{document}